\documentclass[shortnote,twocolumn]{jpsj3}
\usepackage{txfonts}
\usepackage{color}

\title{Machine Learning Phase Diagram in the Half-filled One-dimensional Extended Hubbard Model}

\author{Kazuya Shinjo$^1$, Kakeru Sasaki$^1$, Satoru Hase$^1$, Shigetoshi Sota$^2$, Satoshi Ejima$^{3,4}$, Seiji Yunoki$^{2,4,5}$, and Takami Tohyama$^1$}
\inst{$^1$Department of Applied Physics, Tokyo University of Science, Tokyo 125-8585, Japan \\
$^2$ Computational Materials Science Research Team,
RIKEN Center for Computational Science, Kobe, Hyogo 650-0047, Japan \\
$^3$ Universit\"{a}t Greifswald, 17489 Greifswald, Germany \\
$^4$ Computational Condensed Matter Physics Laboratory,
RIKEN Cluster for Pioneering Research, Wako, Saitama 351-0198, Japan \\
$^5$ Computational Quantum Matter Research Team,
RIKEN Center for Emergent Matter Science, Wako, Saitama 351-0198, Japan
} %\\

\abst{We demonstrate that supervised machine learning (ML) with entanglement spectrum can give useful information for constructing phase diagram in the half-filled one-dimensional extended Hubbard model. Combining ML with infinite-size density-matrix renormalization group, we confirm that bond-order-wave phase remains stable in the thermodynamic limit.}

\begin{document}
\maketitle

The theoretical characterization of phase transitions in strongly correlated systems is important for constructing a phase diagram.
We usually find phase boundaries by detecting anomalous behaviors of physical quantities such as energy.
However, the quantities sometimes exhibit little changes at the boundaries.
Entanglement spectrum (ES) consisting of the eigenvalues of the entanglement Hamiltonian has been suggested as an order parameter.
For example, the absence of a gap called Schmidt gap between the low-lying neighboring two levels of ES is characteristic of topologically ordered phase like the Haldane phase\cite{Lepori2013} and  Kitaev-spin-liquid phase\cite{Shinjo2015}.
However, the Schmidt gap is not necessarily a good quantity for detecting phase transition between non-topological phases.
Another strategy will be the use of machine learning (ML).
In fact, ML has been successful in characterizing ordered states~\cite{Carrasquilla2017}, topological states~\cite{Ohtsuki2016, Nieuwenburg2017}, and photoexcited states.~\cite{Shinjo2019}

In this paper, we show an example of determining a phase diagram in a strongly correlated electron system with the help of  supervised ML. We take the half-filled one-dimensional extended Hubbard model (1DEHM), whose phase diagram has been discussed in connection with the entanglement of the ground-state wavefunctions~\cite{Gu2004,Deng2006,Anfossi2007,Mund2009,Iemini2015,Yu2016}.  
In our ML, ES is used as a training dataset for possible ground states, i.e., the Mott-insulating (MI), charge-density-wave (CDW), bond-order-wave (BOW)~\cite{Nakamura1999}, phase-separated (PS), and superconducting (SC) states.
Using a trained neural network, we determine phase boundaries between the phases in the 1DEHM.

We define the 1DEHM with nearest-neighbor hopping $t$, on-site (nearest-neighbor) Coulomb interaction $U$ ($V$) as
\begin{align}
\mathcal{H}=&-t\sum_{i,\sigma} (c_{i,\sigma}^\dag c_{i+1,\sigma}+H.c.)
+ U\sum_i n_{i,\uparrow}n_{i,\downarrow} + V\sum_i n_{i}n_{i+1}, \nonumber
\end{align}
where the operator $c^\dagger_{i,\sigma}$ is the creation operator of an electron with spin $\sigma (= \uparrow, \downarrow)$ at site $i$ and $n_i=\sum_\sigma n_{i,\sigma}$ with $n_{i,\sigma}=c^\dagger_{i,\sigma}c_{i,\sigma}$. 
We consider a half-filled system and take $t=1$.

Since ES is useful for characterizing phases of 1DEHM~\cite{Gu2004,Deng2006,Anfossi2007,Mund2009,Iemini2015,Yu2016}, we use ES as a training dataset in the present study.
In order to calculate ES, we use infinite-size density-matrix renormalization group (iDMRG) method~\cite{McCulloch2008} .
We keep 400 to 600 density-matrix eigenstates.
In a system composed of two subsystems, A and B, a Schmidt decomposition of a many-body state $|\psi \rangle$ reads $|\psi \rangle = \sum _i p_i |\psi^i_A \rangle |\psi ^i _B \rangle =\sum _i e^{-\xi_i} |\psi _A^i \rangle |\psi _B^i \rangle$,
where $p_i$ is the eigenvalue of reduced density matrix $\rho _A ={\rm Tr}_B |\psi \rangle \langle \psi | =e^{-\mathcal{H}_E}$ for subsystem A and $\xi_i$ is the eigenvalue of the entanglement Hamiltonian $\mathcal{H}_E$ that constructs ES.
We take the subsystem A to be half of the whole system throughout this paper.

\begin{figure}[t]
  \centering
    \includegraphics[clip, width=20pc]{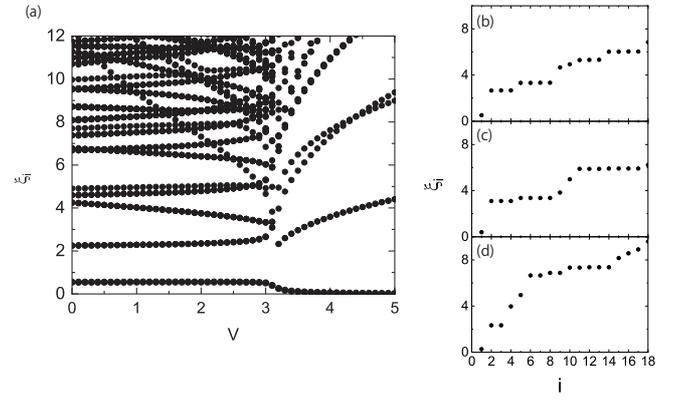}
    \caption{ES for the half-filled 1DEHM obtained by iDMRG. We take $U=6$. (a) $V$ dependence of $\xi_i$. The distribution of $\xi_i$ for (b) $V=2.9$, (c) $V=3$, and (d) $V=3.1$.}
    \label{f1}
\end{figure}

Figure~\ref{f1}(a) shows $V$ dependence of $\xi_i$ for the ground state with $U=6$.
It is known that there are three phases at $U=6$:~\cite{Ejima2007} MI in $0<V\lesssim 3$, BOW in $3\lesssim V\lesssim 3.1$, and CDW in $3.1\lesssim V< 5$.
In Fig.~\ref{f1}(a), we can identify the change of distribution in $\xi_i$ around $V=3$.
In order to make clear phase boundaries around $V=3$, we plot $\xi_i$ for $V=2.9$, 3.0, and 3.1 in Figs.~\ref{f1}(b), \ref{f1}(c), and \ref{f1}(d), respectively. 
We find that the number of degeneracy from the lowest level is $1,3,4,1,1,3,4,\cdots$ for $V=2.9$, $1,3,4,1,1,3,4,\cdots$ for $V=3$, and $1,2,1,1,2,2,2,3,1,1,1,\cdots$ for $V=3.1$.
The sequence of degeneracy does not change between $V=2.9$ and $V=3$, but changes between $V=3$ and $V=3.1$.
In this sense, phase transition at $V\sim 3.1$ between BOW and CDW is easily found by using the degree of degeneracy equivalent to the Schmidt gap.
On the other hand, for phase transition between BOW and MI at $V\sim 3$, there is no obvious signal indicating a qualitative change in ES. 
This may be due to the fact that the phase boundary is of Kosterlitz-Thouless type. 
One of the useful strategies for detecting the change in ES will be the use of ML, which can automatically detect the intrinsic pattern of ES for each phase.

\begin{figure}[t]
  \centering
    \includegraphics[clip, width=20pc]{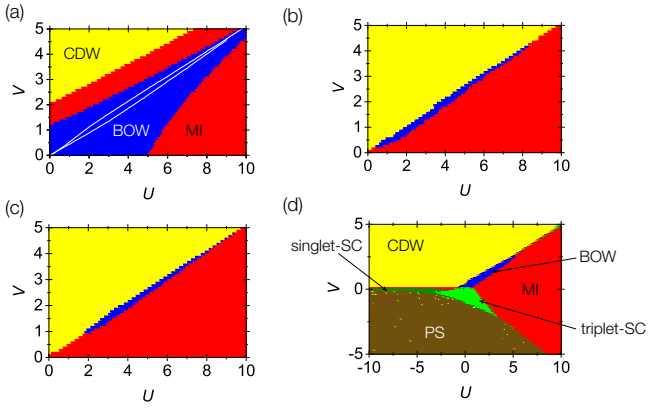}
    \caption{(Color online) Ground-state phase diagram of the half-filled 1DEHM constructed by using ML techniques combined with ES. (a) $K$-means clustering for $U>0$ and $V>0$. The white lines denote the phase boundaries determined by finite-size DMRG~\cite{Ejima2007}. (b) Neural network for $U>0$ and $V>0$. (c) The same as (b) but using ES from finite-size DMRG with $L=120$. (d) Neural network for $-10<U<10$ and $-5<V<5$ using ES from finite-size DMRG with $L=40$. MI, CDW, BOW, PS, singlet-SC, and triplet-SC phases are shown by red, yellow, blue, brown, green, and light green colors, respectively.}
    \label{f2}
\end{figure}

Among various techniques in ML, $K$-means clustering~\cite{MacQueen1967}  is a type of unsupervised learning, which is used when we have data without defined categories or groups. 
This algorithm can find $K$ groups in the data by minimizing the within-cluster sum of squares.
In Fig.~\ref{f2} (a), we show the phase diagram for the parameter space $\{0<U<10, 0<V<5 \}$ obtained by the $K$-means clustering on the ES data with $K=3$ corresponding to CDW, BOW, and MI.
We cannot obtain correct phase boundaries denoted by white lines.
This indicates that the clustering structure of ES is insufficient for constructing phase diagram.

%cannot without supervising.

We next try supervised ML.
Training data are extracted randomly from the following parameter regions: $\{1<U<3,0<V<0.2\}$ and $\{3<U<7,0<V<1\}$ for MI, $\{0<U<0.8,0.5<V<4\}$ for CDW, and $\{2<U<6,V=U/2\}$ for BOW.
The total number of the data for each phase is 10,000.
Using these dataset, we construct a three-layer neural network with one hidden layer, where 200 input units for the 200 lowest ES, 300 hidden units for the hidden layer, and 3 output units to distinguish the CDW, BOW, and SDW phases.
Our network is trained and optimized with the help of Chainer framework~\cite{Chainer}.
The network produces a phase diagram shown in Fig.~\ref{f2}(b), where phase boundaries are consistent with the reported ones~\cite{Ejima2007} denoted as the white lines in Fig.~\ref{f2}(a).

The presence of BOW has been confirmed only for finite-size system up to the system size $L=512$.~\cite{Ejima2007}
On the other hand, our ML result in Fig.~\ref{f2}(b) corresponds to the thermodynamic limit because of the use of iDMRG.
This confirms the stability of BOW in the thermodynamic limit.
Phase boundaries can be determined by order parameters and excitation gaps, while these simulations are performed very naively.
The ML-approach in combination with iDMRG gives an alternative opportunity to obtain phase diagram directly in the thermodynamic limit. 
%Although we can find phase boundaries by calculating order parameters with iDMRG, the ML approach gives an alternative method for obtaining phase diagram in the thermodynamic limit.

In order to know whether the same phase diagram is obtained for finite-size systems, we calculate ES by finite-size DMRG~\cite{White1993} and train the neural network.
Figure~\ref{f2}(c) shows the phase diagram for the $L=120$ chain.
The phase boundaries similar to the thermodynamic tcase in Fig.~\ref{f2}(b) are obtained. 

Encouraged by this small finite-size effect, we apply our ML procedure to a shorter chain with $L=40$ but examine phase boundaries in the full parameter space with $-10<U<10$ and $-5<V<5$, where six phases are known to exist: MI, CDW, BOW, singlet-SC, triplet-SC, and PS states.
We thus use 6 output units.
Training data are extracted randomly from various parameter regions.~\cite{sample_parameters}
In order to improve the accuracy of ML, the neural network is combined with majority voting, which is called model averaging~\cite{Breiman1994, Goodfellow}.
In addition, we use not $\xi_i$ but $\xi_{i+1}-\xi_i$ as a training dataset.
Figure~\ref{f2}(d) exhibits the obtained  phase diagram, being qualitatively consistent with the previous reports (see, for example, Fig.~1 in Ref.~\citen{Iemini2015}). 
However, there are several problems: the MI phase (red in color) is seen near the narrow region of $U\sim 0$ and $V<0$, and  the PS phase (brown in color) is nonuniform with several dots.  
These problems indicate that the classification of all six phases at once is less accurate as compared to the case with only three phases [Figs.~\ref{f2}(b) and \ref{f2}(c)].
In order to classify these phases more accurately, we need to find out a more efficient way to construct the neural network, which remains as a future work.

To summarize, we have demonstrated that the supervised ML whose neural network is trained by inputting ES can give a  phase diagram of the half-filled 1DEHM, which is consistent with the previously known result.
We have also confirmed that BOW remains stable in the thermodynamic limit.
Furthermore, examining unknown states of matter using this supervised ML will be interesting.
One example is an application for photoexcited states in the 1DEHM~\cite{Shinjo2019}. 

\begin{acknowledgment}
%\acknowledgment
This work was supported by CREST (Grant No. JPMJCR1661), the Japan Science and Technology Agency, the creation of new functional devices and high-performance materials to support next-generation industries (CDMSI) to be tackled by using a post-K computer. S. S. was also supported by a Grants-in-Aids for Young Scientists (B) (No. 17K14148) from MEXT, Japan. The iDMRG simulations were performed using the ITensor library.~\cite{ITensor}
\end{acknowledgment}


\begin{thebibliography}{99}
\bibitem{sample_parameters} We take the following parameter spaces as training dataset: $\{5<U<7, 1<V<2\}$, $\{1<U<7, V=U/2-0.3\}$, $\{3<U<4, -1.5<V<-1.0\}$, and $\{4<U<10, -1<V<0\}$ for MI phase, $\{4.5<U<5.5, 4<V<5\}$, $\{-10<U<-5, 0.1<V<1.1\}$, $\{-6<U<-1, 0.01<V<1.01\}$, and $\{-1<U<0, 4<V<5\}$ for CDW phase, $\{0.5<U<1, V=U/2\}$ and $\{3.5<U<4.5, V=U/2\}$ for BOW phase, $\{-3<U<-4, -0.3<V<-0.2\}$ and $\{-9<U<-8, -0.1<V<-0.2\}$ for singlet-SC phase, $\{-2.5<U<-2.0, -1<V<-0.8\}$, $\{0.5<U<0.8, -0.9<V<-0.8\}$, and $\{0.2<U<0.5, -0.7<V<-0.5\}$ for triplet-SC phase, and $\{0<U<1, -4<V<-5\}$ and $\{-4<U<-3, -5<V<-2\}$ for PS phase.
\end{thebibliography}
\end{document}